\def\beq{\begin{equation}}
\def\enq{\end{equation}}
\def\bea{\begin{eqnarray}}
\def\ena{\end{eqnarray}}%
\documentstyle[multicol,aps,prl,psfig]{revtex}
\input{epsf.tex}

\begin{document}

\title{Collapse of Flexible Polyelectrolytes in Multivalent Salt
        Solutions}
\author{ Francisco J. Solis and Monica Olvera de
la Cruz}
\address{Department of Materials Science and Engineering, \\
	 Northwestern University, Evanston, Illinois 60208-3108.}
\maketitle

\begin{abstract}
The collapse of flexible polyelectrolytes in a solution of
multivalent counterions is studied by means of a two state model. 
The states correspond to rod-like and spherically collapsed
conformations respectively. We focus on the very dilute monomer
concentration regime where 
the collapse transition is found to occur when the 
charge of the multivalent salt is comparable (but smaller) to
that of the monomers. The main contribution to the free energy of 
the collapsed 
conformation is linear in the number of monomers $N$, since the internal 
state of the collapsed polymer approaches that of an amorphous ionic
solid. The free energy of the rod-like state grows as $N\ln N$, 
due to the electrostatic energy associated with that shape. 
We show that practically all multivalent counterions added to the
system are condensed into the polymer chain, even before the collapse. 
\end{abstract}

\begin{multicols}{2}
\section{Introduction}

	Nucleic  acids and many proteins are charged and exhibit 
self-assembly.  For
example, DNA with an extended length of the order of centimeters is 
highly compacted in cells and viruses. The compactness of strongly 
charged chains in low ionic
strength solutions has been a subject of recent great interest
\cite{gonzalez,olvera,bloomfield,liu,gelbart,raspaud,arenzon,shklovskii,solis,brilliantov,pincus,thiru}. 
Strongly charged chains, though in
principle in extended conformations to decrease their electrostatic energy,
 precipitate into highly compact structures when small amounts of high
valence counterions are added to low ionic strength solutions 
\cite{olvera,raspaud,widom}. 
Flexible polymers, including single stranded DNA and many
synthetic polymers such as Polyester Sulphonated, precipitate into dense
amorphous structures \cite{olvera}. Semi-flexible polymers, such as long
double stranded DNA, and rigid rods, such as short DNA fragments, aggregate
into highly compacted ordered structures \cite{widom}.  Multivalent induced
DNA compaction is a promising mechanism to ``pack'' DNA, a critical problem in
gene therapy given that long extended DNA cannot be deliver into cells.
Furthermore, the precipitation is correlated with highly accelerated rates
of DNA renaturation and cyclization \cite{siko1,siko2},
and other important biotechnological 
processes. Aside from the DNA compaction and its importance in biology and
biotechnology, the physics of  self-assembly of charged molecules remains an 
interesting open problem. 
	
The self-assembly of polyelectrolytes is strongly related to the
 ``condensation'' of the
counterions along the chains. A fraction of the counterions will be
condensed, strongly attached to the backbone of the polyelectrolytes
to decrease their electrostatic
energy.  In flexible chains the interaction between the polyelectrolytes and
the counterions is non-trivial.  The long range electrostatic
interaction that leads to ion condensation is strongly dependent on the
chain conformation; more ions are condensed when the chain is compacted
than when it is extended.  The conformation of the chain,
reciprocally, depends on the number of condensed counterions,
and their distribution along the chain. The
correlations between monomers and counterions are such that
the higher the number of condensed counterions, the smaller the size of the
chain \cite{gonzalez}. This mechanism induces a transition between two
possible polyelectrolyte conformations: a highly compact shape with nearly
zero effective charge (it is neutralized by the condensed counterions) and
a rod like shape with a smaller number of condensed counterions.  

	The above counterion mediated precipitation mechanism is quite general.
However, it is non trivial to compute the free energy of the
collapsed chains. In this paper we use simple physical arguments and known
results from related charged systems to construct the free energy of the
collapsed state of flexible polyelectrolytes.  We aim to describe the
multivalent induced precipitation in highly dilute polyelectrolyte solutions
at low ionic strength solutions of monovalent ions.  It is well documented
experimentally \cite{olvera,raspaud} that in this regime the
polyelectrolytes precipitate when the concentration of high valence ions is
proportional to the concentration of charged monomers and that the transition 
is quite universal in this regime as it is  independent of the charge
density of the chains and on the specific nature of the condensing
 agents. 
In this paper we recover this result with a first order transition
 model from rod-like to collapsed states described below. We find that at the 
precipitation point all the multivalent ions replace the condensed
 monovalent ions, as expected, and
that the effective charge of the finite size precipitate is nearly zero.


\section{Modeling Considerations}

In our approach we assume that there are two well defined types of
states allowed for the single chain: collapsed and extended. We calculate 
free energies for each of these types of states and study the conditions 
under which one of the states is preferred. Ideally, we would like to present
a single free energy functional valid for the whole range of conformations
of the chain. It is, however, very hard to construct such a functional.
On the other hand, as we aim to show below, there are very distinct
characteristics in the two phases so that treating them as
different objects is not only possible, but provides a clear view of the
reasons behind the transition from extended
to collapsed conformations.

It has long been understood how the extended states arise. The bare charge
of a polyelectrolyte is always reduced by the condensed counterions, but 
it is nevertheless sufficient to create repulsions between the segments
of the chains that overcome the entropic elasticity of the chain, leading 
to an almost fully stretched conformation. 
The collapsed state is, comparatively, much harder to understand. The 
repulsive interaction between the uncompensated charges is still present, 
and therefore, a sufficiently strong cohesive energy is necessary to
maintain the collapsed state. This cohesive energy has been modeled
in previous works by means of virial approximations \cite{gonzalez,raspaud}
and other averages
over positional fluctuations of the charged particles in the system
\cite{liu,brilliantov,pincus}. 
We propose here  
a different approach to the calculation of the 
cohesive energy for the collapsed state .

In several treatments of the problem of attractions between rod-like
polyelectrolytes the attractive energies have been identified as arising 
from strong correlations between the condensed counterions. These 
correlations can be so strong as to induce
the formation of crystalline order of the condensed counterions. In the same
way, we argue that also in the case of flexible polyelectrolytes, purely
electrostatic interactions can lead to the collapse of the  chains. 
A second degree of difficulty arises from the fact that even after 
identifying the origin of the attractive interactions, the calculation 
of the precise value of the energy of interaction can become a technically 
difficult problem. In this paper we will adopt the view that in the 
collapsed state, the monomers and condensed counterions of a
sufficiently large polymer approach a homogeneous bulk state inside
the collapsed region. 
There, we can then estimate the interaction energies 
by considering a similarly composed amorphous ionic solid. One lesson
learned from the study of association of rigid polyelectrolytes is
that it is possible to estimate interaction energies by simply using 
low temperature limits \cite{arenzon,shklovskii,arenzon}. That is, 
within the collapsed region, the entropic considerations produce only
small corrections. 

We can indeed consider a thought-experiment in which we cut the polymer
bonds so as to have a mixture of single-charge monomers and  counterions 
in solution. In this case we expect to observe the typical behavior 
of a salt. When the amount of solvent is large, all particles are 
dissociated, and as the solvent amount decreases, we can observe the
formation of salt crystals. This process occurs at common temperatures and
concentrations.
Rejoining the monomers to form  again the polymer chain modifies the
previous picture in two ways. First, the dissociation in the presence of large 
amounts of solvent can never be complete as the counterions can be freed 
but not the monomeric charges. Second, in the collapsed state,
at low solvent concentrations, the crystallization is hindered by the 
connectivity of the chain, which provides an entropic barrier  to full 
crystallization and imposes a clear, well defined size for the
minimum size of the crystal grain (the collapsed radius of gyration). 
 

On the basis of these considerations we obtain a model free
energy for both states, and calculate the amount of added multivalent
salt required to collapse the chain. While the chain is expected to 
collapse even without the addition of multivalent salt, such collapse
would occur only at very low temperatures, and much of the interest 
on this process is related to applications at room temperatures. 
The reason why the multivalent counterions are so effective at
collapsing the chain also becomes apparent in the elucidation  of the
cohesive energy of the collapsed chain. Since this energy is dominated
by the electrostatic interaction between particles, the replacement
of mono-valent by multivalent charges has the effect of highly 
decreasing the self-energy part of the counterions. This is further 
helped by the fact that it is entropically much more favorable to free
multivalent instead of monovalent counterions. As we will show below, 
this replacement of charges is almost complete even in the
non-collapsed state. 

\section{Free energy calculations}

We describe now the basic parameters of the system. We consider monodisperse
polyelectrolyte chains with monomer number $N$ that we take to coincide 
with the number of monovalent, positive charges in the chain. We assume 
that the distance along the chain  between charges is a constant $b$ that 
is comparable with all other small distances in the system 
such as the thickness (diameter) of the chain and the non-electrostatic
persistence length.  The monomer concentration is $\phi$, and the solution 
contains  several other charged components: a monovalent salt with 
concentration $s$, and a multivalent salt with concentration $m$,
and valence $z$.
We assume that the monovalent counterions naturally associated to the chain
are identical with the  dissociated negative components of the monovalent 
salt. The concentration of these negative ions is  then $s+\phi$. The size 
of all the small particles is also taken to be order $b$. The dimensionless 
Bjerrum number is the ratio $Bj=e^2/4\pi\epsilon b k_{B}T$, with $e$ the
electron charge, $\epsilon$ the permitivity of water, $k_{B}$ the Boltzmann 
constant, and $T$ the temperature.


\subsection{Energy of free charges}

Whether the chain is in a collapsed or in an extended state, the charges
that are not condensed to  the chain give rise to similar contribution 
to the free energy mostly arising form their translational
entropy, and we will consider them all at once. A fraction $f$ of the total
charge in every chain is compensated by condensed counterions, that 
remain in close proximity to the chain. This  compensating charge is made 
up from monovalent and multivalent counterions. The respective
charge fractions that they compensate are $f_{s}$, and $f_{m}$, so that
$f=f_{s}+f_{m}$.  Per chain, we have $f_{s}N$,  and $f_{m}N/ z$, condensed 
negative monovalent and multivalent ions per chain. The free positive ions per
chain of the two different species are simply $sN/\phi$, and $mN/\phi$,
since none of the positive ions condense to the chain. Finally, we have
$(1-f_{s})N+s/\phi$ free negative monovalent ions and $mN/\phi-f_{m}N/z$  
free negative multivalent ions.


For sufficiently dilute concentrations of polymers the electrostatic
contribution of the free charges can be shown to be very small. In a 
Debye-Huckel approximation, the electrostatic energy (in a per chain basis) 
is proportional to $\phi^{1/2}$. Instead, the  main entropic 
contribution is proportional to $\ln \phi$. The volume occupied
by the chain is considered negligible and we obtain for the translational
entropy of the free ions, per monomer, in units of $k_{B}T$:
\bea
        F_{s}&=&  \frac{s}{\phi} \ln {b^3s}
	+ (\frac{m}{\phi}-f_{m}/z)\ln {b^{3}(m-f_{m}\phi/z)} \nonumber \\
      && + \frac{m}{\phi} \ln {b^{3}m} 
       + (1-f_{s}) \ln {b^{3}(s+(1-f_{s})\phi )}.
\ena
In this expression we have estimated the entropy by counting the number of
sites available to  the ions in a lattice with sites of size $b$ that are 
always occupied by ions or solvent molecules.  Since the system is dilute 
we have not included the terms of the form $\l (1-s)b^3$ etc.


\subsection{The Extended State}

        To estimate the electrostatic free energy for the extended state,
we assume a simple linear geometry for the chain. We assume that the chain 
is strongly stretched so that we take the linear size to be simply $L=Nb$. 
The elastic entropy  of the chain plays only a small role, and reduces the
effective size only by a small fraction. Assuming a uniform distribution 
of condensed charges along the chain, the effective charge per monomer 
is $1-f_{s}-f_{m}$. The main contribution to the electrostatic free
energy arises from the interaction of the charges with the logarithmic 
potential created by the effective charge,
\beq
        F_{ex}= Bj (1-f_{s}-f_{m})^2\ln N.
\enq

        For the purpose of comparison with the collapsed state, we note
that this energy is quadratic in the fraction condensed, and  
dependent on the size of the system.

\subsection{The collapsed state}

        We begin with the assumption that the collapsed state is 
very dense. Taking into account the space needed to accommodate the
condensed counterions, we expect the radius of the collapsed polymer 
to be roughly $R=(2Nb^{3})^{1/3}$. To calculate the electrostatic free 
energy of the system we need to determine the distribution of charges 
within the collapsed polymer. To model this distribution we borrow two 
arguments from  classical electrodynamics and solid state physics.


        First, we consider the distribution of uncompensated charges. 
In the classical ({\it i.e.} non-statistical) analog of this problem,
whenever there are free mobile charges within a solid of fixed shape,
the excess charge resides in the surface of the object. 
Thus we can argue that the counterions in the collapsed polymer occupy 
places ``deep'', in the collapsed region to cancel the monomeric charge 
there, while the few uncompensated monomeric charges will appear at 
the surface of the ball. The thermal fluctuations of the system decrease
the sharpness of this result, and we expect a broadening of the regions
in which there can be detected a non-zero average  charge. We separate
the collapsed ball into these two regions, the charged surface and the 
neutral core. Assuming a spherical geometry for the collapsed region, 
the interaction of the charges at the surface is readily calculated by 
an application of  Gauss' law, and is simply
\beq
        F_{es}= Bj (1-f_{s}-f_{e})^2 \frac{N}{(2N)^{1/3}b}.
\enq
As we shall see later, the fraction condensed for the collapsed
state is very close to unity, and the energetic contribution to this 
charged shell is very small.


        Let us consider now the self-energy of the core, which is,
in the average, electrically neutral. When the polymer is collapsed, 
all charges are very close to each other, and their electrostatic 
interactions are very strong. For example, for the case of the aggregation 
of rod-like polyelectrolytes, several model calculations have shown that 
there are very strong correlations between the condensed counterions 
leading to the formation of Wigner crystals along the rods. In the present 
case, however, we cannot argue for such crystallization  because of the 
constraints that the connectivity of the chain imposes. Instead,
we propose to view the core as  an amorphous ionic solid. Finally, 
while we do not expect to find a crystalline state, and only observe 
very strong correlations, the evaluation of the free energy can be 
carried out in a very similar manner to ionic crystals.
This amorphous ionic solid state has also been proposed recently by 
Lee and Thirumalai that they call Wigner Glass \cite{thiru}.


        In the core there is a very strong energetic penalty for 
conformations in which the charge of a particle is not quickly compensated  
by its neighbors. This is clearly the case as well in fully crystalline 
structures, but this requirement is essentially all that is needed for 
a simple estimation of the electrostatic bulk energy of crystal, and 
therefore is also applicable to our case. The energy per atom in an ionic
crystal is usually presented in the following form:
\beq
        E_{eb}=-e^{2}\frac{M}{r_{o}},
\enq
where $r_{o}$ is the lattice constant and $M$ is the Madelung 
constant \cite{feynman}. 
The Madelung constant depends on the specific lattice structure, but its
value is always of order $1$. The 
sum of the interaction of one atom with all other atoms in the 
crystal is arranged in a convergent series of alternating terms, 
summing the interaction with a layer of oppositely charged atoms,
(nearest neighbors) then with the layer of next-nearest neighbors that
are equally charged, etc. Clearly, the final result is dominated 
by the interaction with the neighbors that compensate the charge 
of a given particle, and the rest simply modify the prefactor. In the 
following we sketch a simple estimate for an effective Madelung constant 
for the collapsed polymer bulk. 


        We have assumed that all small distances in the system are 
of order $b$. We cluster one counterion with all the monomers
that are required to compensate its charge, $1$ for monovalent,
and $z$ for multivalent ions. If all the particles in the cluster are at 
about the same distance from each other, the energy contribution
of the cluster is 
\beq
        \frac{E}{kT}=-Bj\frac{z}{2}(z+1).
\enq  
This is the first term in the series for the effective Madelung constant,
and we will truncate the series at this point. There are $f_{s}N$ clusters
with monovalent counterions, and $f_{m}N/z$ with multivalent counterions,
so that our estimate for the free energy of the bulk of the collapsed
state is 
\beq
        F_{eb}=-Bj(f_{s}+\frac{1}{2}f_{m}(z+1)).
\enq


\section{Phase Transition Conditions.}

We can now use the free energy expressions postulated for the 
collapsed and extended states to calculate the phase diagram
of the system which we shall present in $\phi-m$ coordinates. 
It is useful, however, to carry out further simplifications in
the model that lead to some qualitative and quantitative
results. 


\subsection{Replacement by multivalent counterions}

The general expressions for the free energy in the two states,
collapsed:
\beq
	F_{c}=F_{eb}+F_{s},
\enq
and extended:
\beq
	F_{x}=F_{ex}+F_{s},
\enq
still depend
on the fractions condensed $f_{s}, f_{m}$. These fractions
should be obtained before computing the final free energy. 
To obtain their values, the free energy is minimized with 
respect to these variables, which leads
to the equations:
\bea
        2Bj (1-f_{m}-f_{s})\ln N 
        + \ln {b^{3}(s+(1-f_{s})\phi )} =0, \nonumber \\ 
        2Bj (1-f_{m}-f_{s})\ln N 
        + (1/z) \ln {b^{3}(m-f_{m}\phi/z)}=0. \label{fminsr}
\ena
for the extended state, while for the collapsed state one has:
\bea
        &&\frac{Bj N}{N^{1/3}}(1-f_{s}-f_{m})+
        Bj  + \ln {b^{3}(s+(1-f_{s})\phi )} =0, \nonumber \\ 
        &&\frac{Bj N}{N^{1/3}}(1-f_{s}-f_{m})+
        Bj \frac{z+1}{2} + 
	\frac{1}{z} \ln {b^{3}(m-\frac{f_{m}\phi}{z})}=0. \label{fminbu}
\ena


The Eqs.~(\ref{fminsr}) imply a particular relation between 
the fraction condensed of the two species, regardless of the strength
of the electrostatic term (measured by $Bj$), given by
\beq
        (b^{3}(s+(1-f_{s})\phi )^z=b^{3}(m-f_{m}\phi/z). \label{replace}
\enq
This relation is essentially a result of the fact the same
energy binding $z$ monovalent counterions is what is required to bind
one multivalent ion, while the entropy gain of freeing $z$ ions is
also $z$ times bigger. For dilute concentrations of monomers and 
ions, the expressions in Eq.~(\ref{replace}) are all small quantities and 
even for $z=2$ the exponentiation of the left hand side of the equation 
leads to an extremely small value for the difference $m-f_{m}\phi/z$.
Thus we conclude that for all practical purposes {\it all multivalent
counterions are condensed}. At higher concentration of these counterions,
there can be a near saturation since the polymer cannot absorb
all of them, and it cannot be emptied of monovalent ions (since by 
Eq.~(\ref{replace}) $f_{s}$ cannot be zero.)


For the second pair of equations Eq.~\ref{fminbu} we note that  
the first terms of the equations, which arise from the electrostatic 
repulsion of uncompensated charges, present very large prefactors
to the effective charge $1-f_{s}-f_{m}$. Clearly, the effective 
charge must cancel this contribution, and we conclude that,
essentially, in the collapsed state {\it all charges are compensated}. 
An analysis similar to the case of the extended conformation implies
that the charge compensation must occur first with as many multivalent
ions as possible, and then finalize with monovalents so that, roughly, 
we have $f_{m}=zm/\phi$, and $f_{s}=1-f_{m}$.


\subsection{Low Density Limit}

Let us now consider the case in which the density of monomers is
very low. More precisely
we define this regime by the condition that if all chains are
extended, they do not overlap. Since the extended size of a chain is 
$Nb$, the definition is equivalent to $\phi < N/(Nb)^{3}=1/N^2b^3$. 


To obtain a simplified formula for the amount of multivalent 
counterions needed to produce the transition, we need to 
carry out some simplifications. 
As the solution becomes more and more dilute, there is a great
entropic energy to be gained by freeing the monovalent condensed
ions. At the same time, the relation obtained between the amount
of condensed ions of the two species still implies that a large 
number of multivalent counterions are condensed. Thus, a first
simplification consist on setting, for the extended state 
$f_s=0$, and $f_{m}=zm/\phi$. Thus, neglecting some 
other small terms, the energy for the extended state reads:
\beq
        F_{x}= Bj(1-\frac{zm}{\phi})^2\ln N
        -(1+\frac{s}{\phi})\ln b^3(\phi+s).
        \label{redrod}
\enq
In the collapsed state, we have already discussed that at all
concentrations we have almost exactly $f_{m}=mz/\phi$, and 
$f_{s}=1-f_{m}$, which leads to 
\beq
        F_{c}= -Bj(1-\frac{z(z-1)m}{2\phi})
        -\frac{zm}{\phi}\ln b^{3}(\phi +s). \label{rebluk}
\enq
In the evaluation of the entropic term we have used the fact that the 
density of required counterions is of the order of the  monomer
density, so that $\ln b^3 f_s(\phi+s) \approx \ln b^3 (\phi +s)$.


Equating the previous two approximations for the free energies in the
two states leads to a quadratic equation in $zm$ for the density of
multivalent counterions required to create the transition. However, 
from the analysis of these equations, a simpler conclusion can be drawn. 
By looking to just the entropic type terms, of the form
$\ln{b^{3}(\phi+s)}$, which are the largest
contributions to these expressions, it is clear that a transition is 
only possible when the charge of the multivalent particles approaches
that of the monomers:
\beq
	zm \sim \phi. \label{punch}
\enq  
This is the resulting limiting condition for very dilute solutions. 
We should remark that while this result might appear intuitive, 
it relies on the specific structure of the extended and collapsed 
state. Namely, in both states, all multivalent ions are trapped, 
in the collapses state, the only free monovalent counterions 
are those that have been replaced by multivalent ones, and 
in the extended state, all monovalent counterions are free. 



\subsection{Dense limit}

In the opposite case, for higher concentrations, the model 
we use cannot give a precise answer as we have neglected the
effects of the electrostatic self-energy of the free counterions. 
Further, as the concentration increases it becomes harder 
to distinguish the ``inside'' and ``outside'' regions of the chain. 
We can obtain, however, a clear limit for the validity of these 
approximations, that also signals the onset of new phenomena.
In the absence of multivalent counterions, the fraction of monovalent
counterions condensed is given to a good approximation by
\beq
	f_{s}=1+\frac{\ln b^{3}(\phi+s)}{2 Bj \ln N}.
\enq
At the onset of the dilute regime $\phi=b^{-3}N^{-2}$, we recover the 
Manning formula $1-f=1/Bj$. Further increase on the concentration leads
to a situation in which the fraction condensed can reach a value 
very close to $1$, so that the effective charge is almost zero and 
cannot extend the chain. After this point the distinction between the 
collapsed and non-collapsed states is much less clearly defined. The 
chain cannot be extended if the electrostatic free energy per monomer
smaller than $1 k_{B}T$ (for flexible polymers). This leads then to
the following criteria: 
\beq
	-\ln b^3(\phi+s) \sim (Bj\ln N)^{1/2}. \label{oneBj}
\enq

\subsection{Monovalent Collapse}

As mentioned in the introduction, it is expected that monovalent 
counterions on themselves can collapse the chain even in dilute
conditions, if the temperature is sufficiently low. Our model gives a
simple criteria for the occurrence of this collapse. 
The free energy for the extended state can be approximated as 
\beq
	F_{x}=-\frac{(\ln b^{3}(\phi+s))^2}{4 Bj \ln N},
\enq
while the energy for the collapsed state is simply approximated by 
\beq
	F_{c}=-Bj. 
\enq
This results in a transition occurring at  
\beq
	-\ln b^3(\phi+s) \sim Bj (\ln N)^{1/2}. \label{twoBj}.
\enq
We remark that this equation differs from Eq.~(\ref{oneBj}), by a 
$Bj^{1/2}$ factor. This difference implies that the 
collapse of the chain by its natural monovalent counterions 
occur within the dilute regime.  


\section{Results}


Solution of the equilibrium equations between the two phases produces
a phase diagram that is valid at low concentrations of monomers. We now 
present results for 
valences $z=2,3$, and $Bj=2$,$3$,$4$,$5$. Other parameters take
values of $N=10^5$, $b^3s=10^{-12}$, while the data covers monomer 
concentrations ranging from $b^3\phi=10^{-18}$, to $10^{-8}$. These results 
are presented in Fig. 1-4. Figures 1 and 3 present results for
the  concentration  $m$, while 2 and 4 show the required ratio
$m/\phi$. 


\begin{center}
\begin{minipage}[H]{3.2in}
\epsfxsize=3.0in \epsfbox{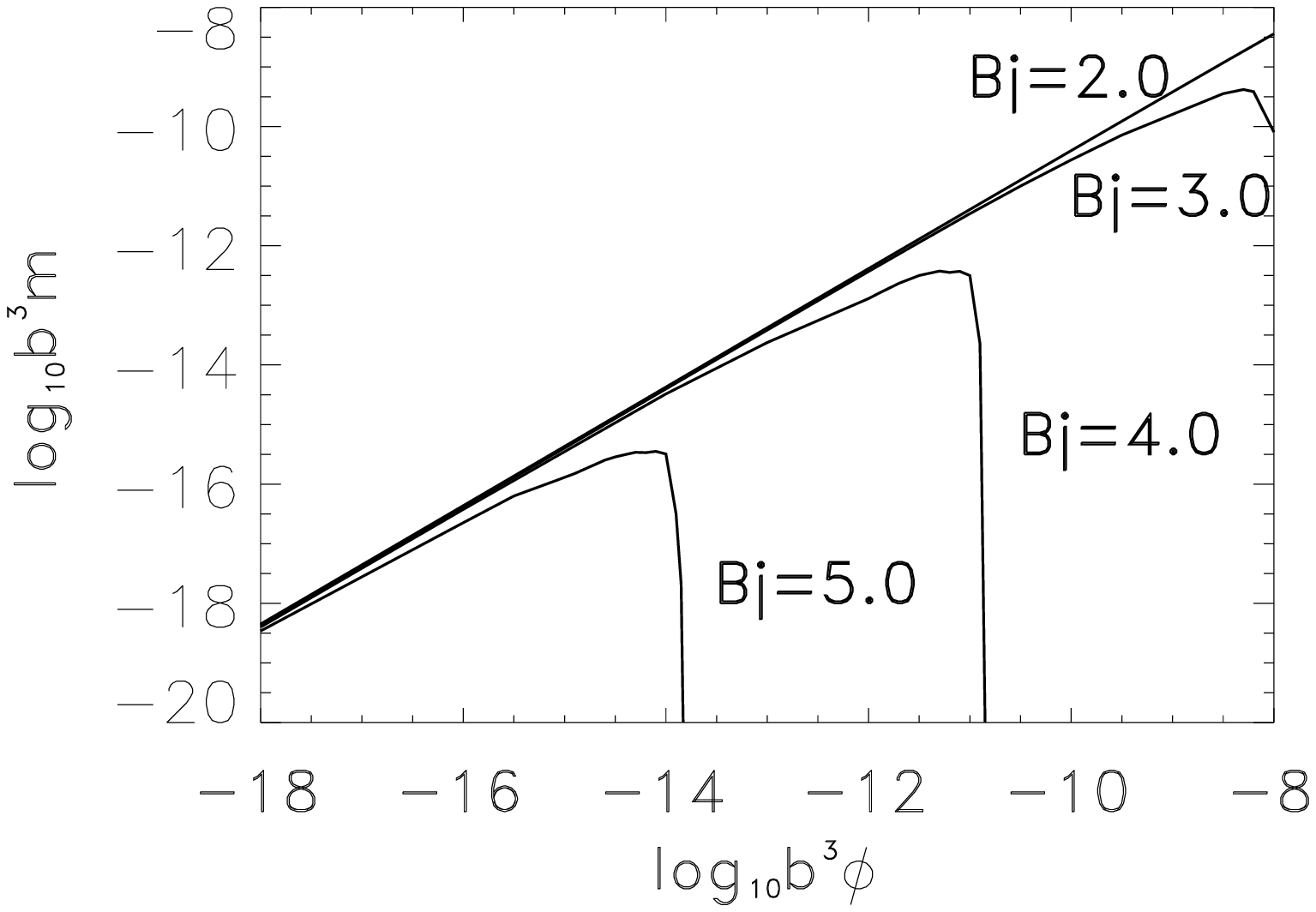}
\begin{figure}
\caption{Plot of $\log b^{3} m$, for z=2 and diferent values of Bj}.
\end{figure}
\end{minipage}
\end{center}


\begin{center}
\begin{minipage}[H]{3.2in}
\epsfxsize=3.0in \epsfbox{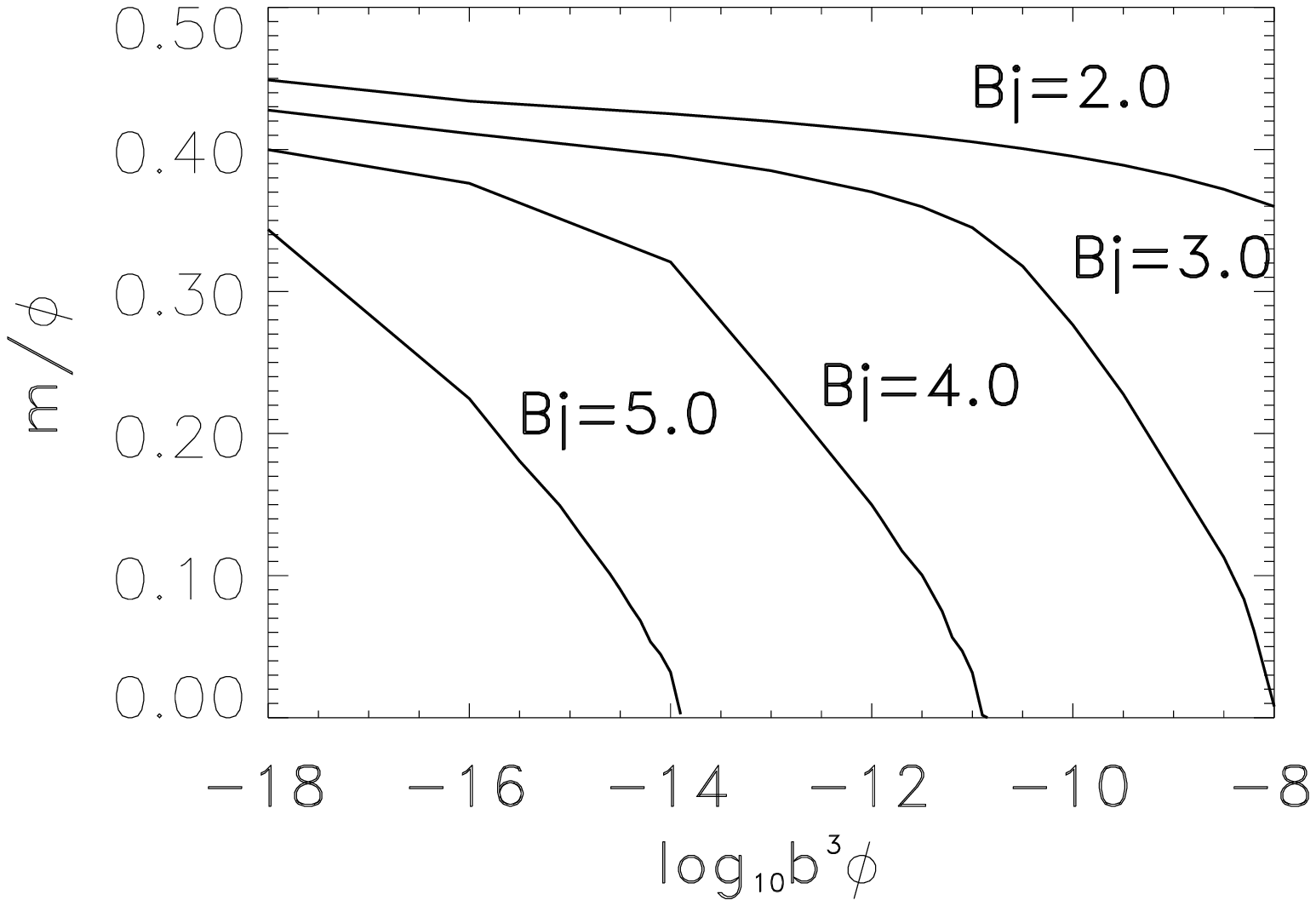}
\begin{figure}
\caption{Plot of the ratio $m/\phi$, for z=2 and diferent values of Bj}.
\end{figure}
\end{minipage}
\end{center}


\begin{center}
\begin{minipage}[H]{3.2in}
\epsfxsize=3.0in \epsfbox{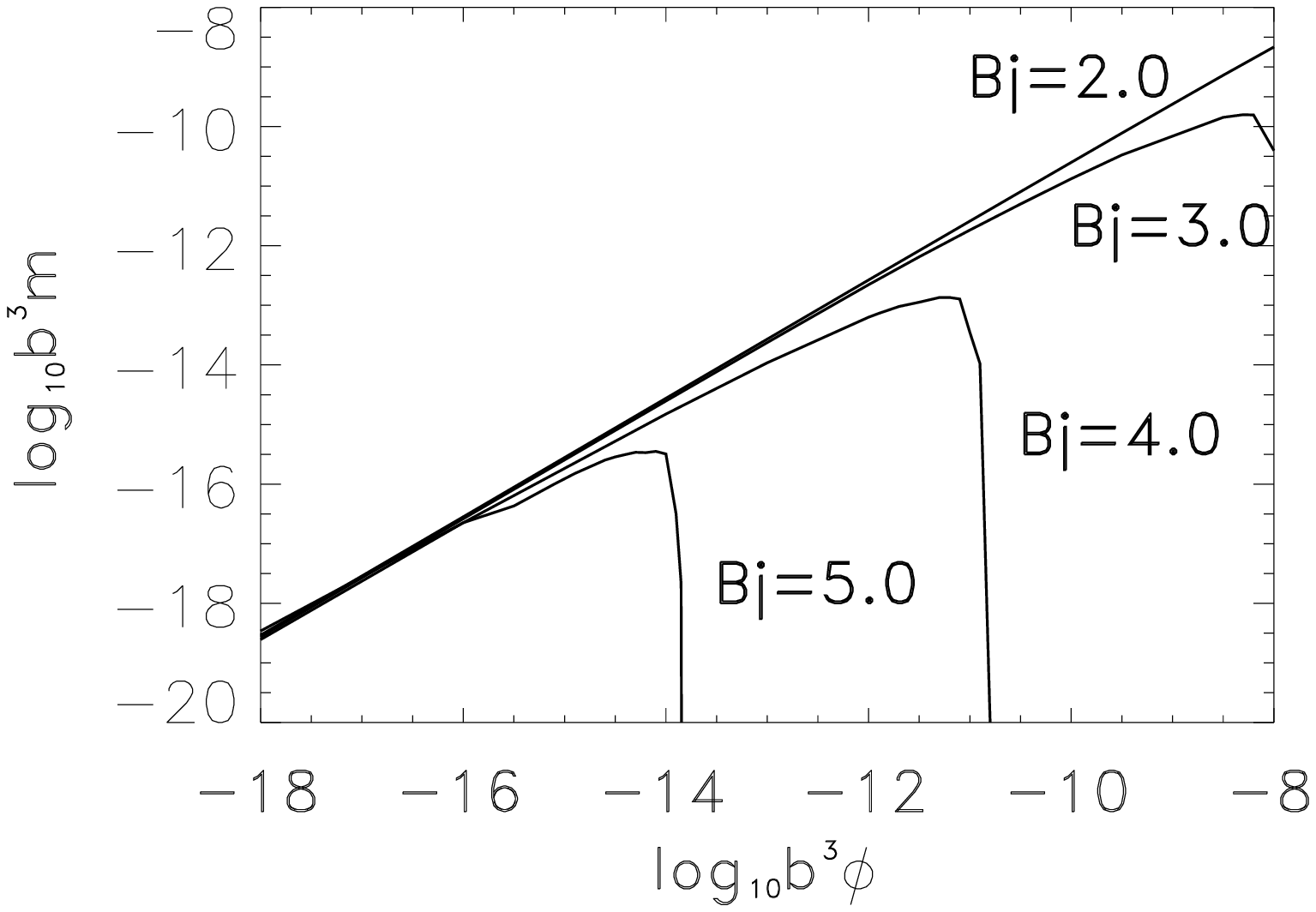}
\begin{figure}
\caption{Plot of $\log b^{3} m$, for z=3 and diferent values of Bj}.
\end{figure}
\end{minipage}
\end{center}


\begin{center}
\begin{minipage}[H]{3.2in}
\epsfxsize=3.0in \epsfbox{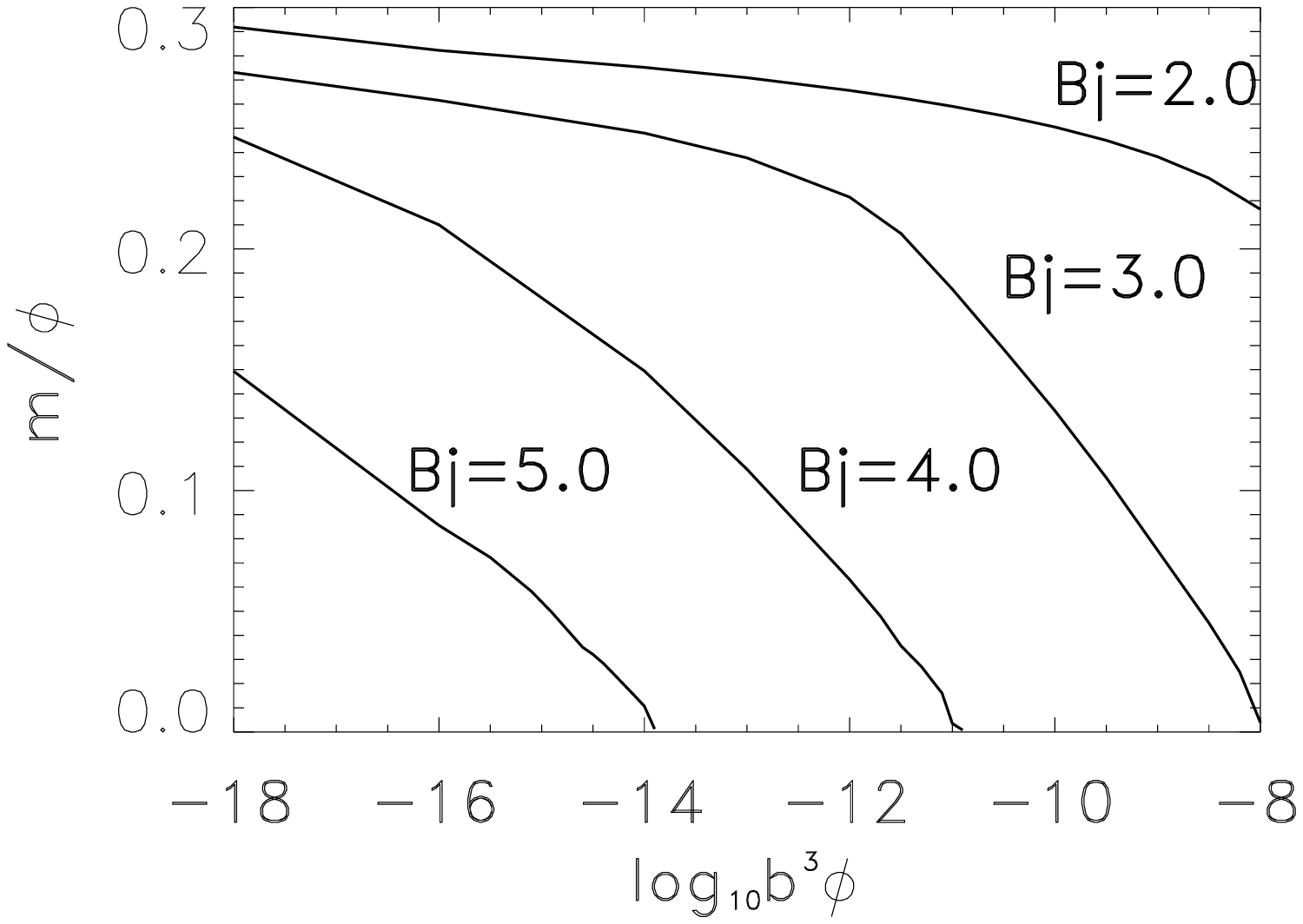}
\begin{figure}
\caption{Plot of the ratio $m/\phi$, for z=3 and diferent values of Bj}.
\end{figure}
\end{minipage}
\end{center}

In both cases, $z=2$, and $z=3$, it is clear that as the density of 
monomer is decreased, the required amount of monovalent counterions 
required for the collapse falls in the range prescribed by the 
limiting relation Eq.~(\ref{punch}), and the ratio is always somewhat smaller
than $1/z$. 

\subsection{Limits of Applicability}

	In this subsections we collect the main limitations of our
approach. The  concentration of multivalent ions at the extended-to-collapse
transition $m^{*}$ appears to be linear with the concentration of charged
monomers $\phi$ at highly dilute solutions, in agreement with experimental
results \cite{raspaud}. As $\phi$ increases, however, $m^{*}$ decreases.
Experimental results indicate the opposite behavior \cite{olvera,raspaud}.
This disagreement has two sources: we are 
neglecting the interchain interactions and also the possibility of
other conformations.  

	In the collapsed state the chains are nearly neutral
so one can neglect the chain-chain interactions provided the concentration
is still much lower than the overlap concentration. However, this is not
the case of other non-collapsed conformations which have higher effective
charge, such as the extended state. 
Our model takes into account
only two states (the stretched and collapsed) of
{\it non-interacting} chains, so that our results are valid only for
highly dilute solution of polyelectrolytes, as has been roughly
established in
Eq.~(\ref{oneBj}). In a more dense regime the electrostatic 
free energy of the free counterions becomes relevant while here
it has been neglected.  
Also, our results are only reliable
in low ionic strength solutions of monovalent ions and at low concentration
of multivalent ions.  When the concentration of multivalent ions increases
there will be an excess of free multivalent counterions so their
electrostatic contribution cannot be neglected.
We have also not taken into account the effects of the 
solvent quality. In poor solvents, for exmaple,  it is know that 
a ``necklace'' structure might be more favorable than the stretched
conformation \cite{kardar,rubinstein,mine}.

\section{Conclusions}

	A collapsed flexible polyelectrolyte in low ionic strength 
solutions has a nearly zero effective charge. The charge of the
monomers is compensated by condensed counterions. The main
contribution to the electrostatic free
energy of the collapsed polyelelctrolyte is proportional to the number of
compensated monomer charges $N$, and the internal state of the
collapsed molecule is similar to an amorphous ionic solid. There is a 
strong electrostatic free energy reduction of the collapsed state when
the  valence of the condensed counterions increases. Therefore, 
when multivalent counterions are added at a concentration $m$ less 
than the concentration of charged monomers $\phi$,
all of the multivalent counterions are condensed, replacing the monovalent
condensed counterions.  Since the multivalent ions reduce the electrostatic
free energy, a first order transition occurs from extended to collapsed chain
conformations.

\subsection*{Acknowledgments}

	This work was sponsored by the National Science Foundation,
grant DMR9807601.

\end{multicols}
\end{document}